\documentclass[aps,prd,showpacs,notitlepage,nofootinbib,superscriptaddress,floatfix,showkeys,twocolumn]{revtex4-2}
\usepackage{blindtext}
\usepackage{hyperref}

\usepackage{amsmath,amssymb}
\usepackage{float}
\usepackage{microtype}

\usepackage{graphicx}
\usepackage{bm}
\usepackage{latexsym}
\usepackage{epsfig}
\usepackage{psfrag}
\usepackage[dvipsnames]{xcolor}
\usepackage{subfigure}
\usepackage{graphicx}

\usepackage{amssymb}
\usepackage{amsmath}
\usepackage{bm}
\usepackage{latexsym}
\usepackage{epsfig}
\usepackage{psfrag}
\usepackage[normalem]{ulem}
\usepackage{textcomp}
\usepackage{color}
\usepackage{pstricks}
\usepackage[utf8]{inputenc}
\usepackage{comment}

\def\nn{\nonumber}
\def\Mpl{M_{_\mathrm{Pl}}}
\def\HI{H_{_\mathrm{I}}}

\def\d{\mathrm{d}}
\def\e{\mathrm{e}}
\def\l{\left}
\def\r{\right}
\def\f{\frac}
\def\ak{\alpha_k}
\def\bk{\beta_k}

\def\ps{\mathcal{P}_{_\mathrm{S}}}
\def\ns{n_{_\mathrm{S}}}
\def\As{A_{_\mathrm{S}}}
\def\dneff{\Delta{N}_{\mathrm{eff}}}
\def\rhore{\rho_{\mathrm{re}}}
\def\Tre{T_{\mathrm{re}}}
\def\Nk{N_{\mathrm{k}}}
\def\Nre{N_{\mathrm{re}}}
\def\ogw{\Omega_{_\mathrm{GW}}}
\def\oR{\Omega_{_\mathrm{R}}}
\def\rhogw{\rho_{_\mathrm{GW}}}

\def\rhore{\rho_{\mathrm{re}}}
\def\ke{k_\mathrm{e}}
\def\kre{k_\mathrm{re}}

\def\phie{\phi_{\mathrm{e}}}
\def\rnbd{r_{_\mathrm{NBD}}}
\def\gk{\gamma_k}
\def\fnl{f_{_\mathrm{NL}}}
\def\wre{w_{\mathrm{re}}}

\def\ka{k_1}
\def\kb{k_2}
\def\kc{k_3}
\def\vka{\bm{k}_1}
\def\vkb{\bm{k}_2}
\def\vkc{\bm{k}_3}

\begin{document}

\title{ACT-ing on inflation: Implications of non Bunch-Davies initial condition and reheating on single-field slow roll models}
\author{Suvashis Maity}
\email{E-mail: saamaity@gmail.com}
\affiliation{Indian Institute of Science Education and Research, Pune 411008, India}

\begin{abstract}
We investigate a class of slow roll inflationary models in light of the recent Cosmic Microwave Background constraints from Planck 2018, ACT DR6, DESI DR1, and BICEP/\textit{Keck} 2018. The combined dataset favors a higher value of the scalar spectral index $n_{_\mathrm{S}}=0.9743 \pm 0.0034$, which places increased pressure on several conventional inflationary scenarios. In this study, we analyze the observational viability of various well-motivated models, including the $\alpha$-attractor E- and T-models, chaotic inflation, hilltop inflation, and natural inflation. 
We incorporate the effects of a post-inflationary phase of reheating and examine how the dynamics of reheating influence the predictions in the $n_{_\mathrm{S}}-r$ plane. We also impose a lower bound on the reheating temperature based on the constraint from the effective number of relativistic species ($\Delta{N}_{\mathrm{eff}}$) arising from primordial gravitational waves. While reheating improves agreement with observations for some models, significant regions of parameter space remain disfavored. 
Finally, we explore the impact of a non Bunch-Davies initial state and demonstrate that it can substantially improve the fit to the $n_{_\mathrm{S}}-r$ data across a broader class of inflationary models, thereby offering a potentially viable mechanism for reconciling theory with the latest observations.
\end{abstract}
\maketitle


\section{Introduction}

One of the key aspects of cosmology is understanding the origin of the tiny anisotropies (of the order of $\sim10^{-5}$) observed in the cosmic microwave background (CMB) and the large-scale structure of the universe.
Over the past decade, these anisotropies have been measured with remarkable precision, placing tight constraints on both their amplitude and spectral index.
Inflation provides a natural mechanism for amplifying quantum vacuum fluctuations, which can give rise to such anisotropies~\cite{Martin:2003bt,Bassett:2005xm,
Baumann:2008bn,Sriramkumar:2012mik}.
In particular, slow roll inflation can generate perturbations with an amplitude and spectral tilt that are consistent with observations.

The Planck 2018 data constrained the scalar spectral index to $\ns=0.9651\pm 0.0044$~\cite{Planck:2018jri,Planck:2018vyg}. 
An improved constraint, combining Planck data with measurements from the South Pole Telescope (SPT), yielded $\ns=0.9647\pm 0.0037$~\cite{SPT-3G:2024atg}.
Recently, the latest data released by the Atacama Cosmology Telescope (ACT)~\cite{ACT:2025fju,ACT:2025tim}, when combined with Planck data, has provided a significantly different constraint of $\ns=0.9709\pm 0.0038$. 
Considering the DESI DR1~\cite{DESI:2024uvr,DESI:2024mwx} (P-ACT-LB-BK18), the constraint has further improved to $\ns=0.9743\pm 0.0034$~\cite{ACT:2025fju}. 
{ Note that the values of $\ns$ are quoted with $68\%$ confidence levels (CL).}
{ This value of $\ns$, that is closer to $1$, which suggests a spectrum that is closer to scale invariance compared to previous results~\cite{Planck:2018jri,Planck:2018vyg,SPT-3G:2024atg,DESI:2024uvr,DESI:2024mwx}.
Further, the abbreviations refer to P (Planck~\cite{Planck:2018jri,Planck:2018vyg}), LB (BAO from DESI DR1~\cite{DESI:2024uvr,DESI:2024mwx}) and BK18 (BICEP/Keck 2018~\cite{BICEP2:2018kqh,BICEP:2021xfz}).
} 
It is also worth noting that the upper bound on the tensor-to-scalar ratio~\footnote{ The tensor-to-scalar ratio, $r$, represents the ratio of the power spectrum of primordial tensor power spectrum to that of scalar power spectrum. 
Although tensor perturbations have not yet been directly observed on large scales, current measurements provide a strong upper bound on $r$, as quoted in the text. 
The current upper bound predicts a low strength of the tensor perturbation. 
Also, such an upper bound on 
$r$ can help rule out certain inflationary models and break the degeneracy among models that predict similar scalar power spectra. }
is chosen $r<0.036$~\cite{BICEP:2021xfz}. 
An updated bound given in~\cite{Tristram:2021tvh}  of $r<0.032$ with $95\%$ CL and P-ACT-LB-BK18 provide $r<0.038$ with $95\%$ CL~\cite{ACT:2025tim}.
In this work, our primary objective is to achieve the desired value of $\ns$. To this end, we explore the parameter space of various quantities that accommodate the required $\ns$ while ensuring that $r$ remains below its observational upper bound. Throughout the analysis, we therefore impose $r<0.036$ as a working constraint.
Further, the bounds on $\ns$ in the $\Lambda$CDM cosmology depend on whether $r$ is allowed to vary or is fixed to zero. Typically, constraints on $\ns$ are slightly tighter when 
$r=0$ compared to the case when $r$ is allowed to vary, because adding $r$ introduces an additional degree of freedom and enlarges the allowed parameter space for $\ns$. We therefore report the bounds on $\ns$ for $r=0$. 
At this point, we should also mention that the value of $\ns=0.9743\pm 0.0034$ that we have used in this work was obtained by taking into account the result of DESI DR1, which assumes $\Lambda$CDM~\cite{ACT:2025fju,ACT:2025tim}.
On the other hand, the result of DESI DR2~\cite{DESI:2025gwf}, strongly disfavors $\Lambda$CDM. Hence, to consider DESI DR2 results, it is advisable to wait for the more recent data release by DESI (for a detailed discussion on this, see~\cite{DESI:2025gwf,Kallosh:2025ijd}).

In~\cite{Forconi:2021que}, the authors varied, in addition to the six standard cosmological parameters, extra parameters such as the running of the scalar spectral index, the tensor-to-scalar ratio, and the spatial curvature. They demonstrated that these additional parameters have only a weak impact on the upper bound of $r$, whereas $\ns$ is significantly affected. In particular, a value of $\ns \sim 0.97$ 
can be obtained using the Planck18 + BK15 data, and an even higher value of $\ns$ is achievable when WMAP data are combined with ACTPol and SPT3G.
The authors of~\cite{Kallosh:2021mnu} demonstrated that the Starobinsky, Higgs, and $\alpha$-attractor models provide a good fit to the Planck18 + BK18 data.  \cite{deHaro:2024gml} discusses that, with suitable reheating parameters, $\ns$ can reach $\sim 0.97$ with $n\to\infty$ in the $\alpha$-attractor model.

The latest constraints (after ACT) on the $\ns-r$ plane have renewed interest in re-evaluating existing inflationary models. It has been shown that while the chaotic inflation model with a $\phi^{1/3}$  potential is now consistent with current data, the Starobinsky $R^2$ model~\cite{Starobinsky:1980te} is nearly pushed outside the $2\sigma$ confidence region~\cite{ACT:2025tim}. Furthermore, inflation with a $m^2\phi^2$ potential can also be brought into agreement with 
observations when a non-minimal coupling to gravity is included~\cite{Kallosh:2025rni}. Other theoretically motivated scenarios, such as Higgs inflation~\cite{Aoki:2025wld,Gialamas:2025kef,McDonald:2025odl,Yin:2025rrs,Yin:2022fgo} and warm inflation~\cite{Berera:2025vsu}, and others~\cite{Dioguardi:2025vci,Salvio:2025izr,Antoniadis:2025pfa,Kim:2025dyi,Dioguardi:2025mpp,Gao:2025onc,He:2025bli,Pallis:2025epn,Drees:2025ngb,Haque:2025uri,Haque:2025uis,Liu:2025qca,Chakraborty:2025yms,Yogesh:2025wak} have also been revisited in light of recent ACT data. Notably, there have been several recent efforts to bring plateau-like models—especially Starobinsky inflation—back into the allowed parameter space, either by incorporating a phase of reheating~\cite{Drees:2025ngb,Haque:2025uis} or by adopting 
non Bunch-Davies (NBD) initial conditions~\cite{Brahma:2025dio}. 

In this paper, we consider a set of single-field slow roll inflationary models that include both power-law and plateau-like potentials, and examine whether they are consistent with the combined P-ACT-LB-BK18 constraints. For models that fall outside the allowed region, we investigate whether they can be brought back into agreement with observations. To this end, we adopt the following strategy:
\begin{itemize}
    \item 
Introduce a subsequent phase of reheating to investigate whether it can shift the scalar spectral index $\ns$
  to the desired range.
\item
Analyse the influence of NBD initial conditions on the $\ns-r$ predictions.
\item
Examine the impact of the $\dneff$ constraint on the primordial gravitational waves generated in these scenarios, and further on the reheating temperature. 
\end{itemize}
We will discuss each of these three aspects in detail in the following sections. Throughout the analysis, we work with a spatially flat FLRW background and assume that the scalar field is minimally coupled to gravity.

The NBD initial conditions refer to excited states of quantum fluctuations that deviate from the standard Bunch–Davies (BD) vacuum, typically characterised by the Bogoliubov transformations of the perturbative modes. Such initial states can naturally arise in scenarios with a pre-inflationary phase~\cite{Borde:2001nh,Kinney:2023urn,Pavlovic:2023mke,Melcher:2023kpd}, trans-Planckian physics~\cite{Martin:2000xs,Kempf:2000ac,Kaloper:2002uj,Hui:2001ce,Schalm:2004qk,Ashoorioon:2004wd}, or deviations from standard slow roll dynamics~\cite{Lello:2013mfa,Jain:2009pm,Ragavendra:2020old}. NBD initial conditions can lead to scale-dependent modifications in the power spectra, thereby altering key observables such as $\ns$, $r$, and non-Gaussianity. Importantly, they can suppress or enhance the tensor-to-scalar ratio, allowing otherwise excluded inflationary models to be consistent with current $\ns-r$ bounds. The predictions of $\ns-r$ can further be modified by the phase of reheating. 

The phase of reheating is the intermediate epoch between the end of inflation and the onset of the standard radiation-dominated era, during which the inflaton undergoes a quasi-periodic oscillation and decays into relativistic particles constituting the standard model (SM). The phase of reheating governs the thermal history of the universe and sets the initial conditions for Big Bang Nucleosynthesis (BBN)~\cite{Kawasaki:1999na,Kawasaki:2000en,Hasegawa:2019jsa}. Reheating is often characterised by parameters such as the reheating temperature $\Tre$, duration, and an effective equation of state (EoS) $\wre$. As stated, the phase influences the mapping between the predictions from inflation and observables in CMB. Specifically, it affects the number of e-folds between the pivot scale and the end of inflation, thereby shifting the prediction of $\ns-r$. 

We also consider an additional constraint arising from the overproduction of primordial gravitational waves (GWs) generated by vacuum fluctuations during inflation. The EoS during reheating can modify the shape of the GW modes that re-enter the Hubble radius during this phase~\cite{Bernal:2019lpc,Haque:2021dha,Maity:2024odg,Maity:2024cpq}. For an EoS with $\wre<1/3$, the spectrum acquires a red tilt, whereas for $\wre>1/3$, the tilt becomes blue. Furthermore, $\Tre$ and $\wre$ dictate the duration of reheating, thereby determining how long the modified tilt persists in the GW spectrum. It is important to note that inside the Hubble radius, GWs behave like radiation. An excess of GWs (which arises for $\wre>1/3$) thus contributes to the effective number of relativistic degrees of freedom, which can be constrained by BBN and CMB~\cite{Jinno:2012xb}. Consequently, these bounds provide a constraint on the duration of reheating or eventually on $\Tre$ for a given $\wre$.

This paper is organized as follows. 
In Sec.~\ref{sec:NBD-state}, we review the framework of slow roll inflation and the generation of primordial perturbations, focusing on the two-point correlations or the power spectrum. We also discuss how a NBD initial state modifies key observables such as the scalar spectral index $\ns$ and the tensor-to-scalar ratio $r$.
In Sec.~\ref{sec:reheating}, we examine the post-inflationary phase of reheating, exploring how the reheating parameters affect the duration of inflation and subsequently alter the predictions for $\ns$ and $r$.
Sec.~\ref{sec:pgw-dneff} is devoted to the impact of $\dneff$ on the amplitude of primordial gravitational waves. We derive the resulting lower bound on the reheating temperature from current observational constraints.
In Sec.~\ref{sec:models}, we analyze a variety of single-field slow roll inflationary models and investigate how the inclusion of the dynamics of reheating and the NBD initial condition influence their compatibility with the latest CMB observations from the P-ACT-LB-BK18 dataset.
Finally, in Sec.~\ref{sec:conclusions}, we present our conclusions along with a brief outlook for future directions.

Let us discuss a few points about the notations. 
We shall work with natural units such that $\hbar=c=1$, and the reduced 
Planck mass is $\Mpl=\l(8\,\pi\, G\r)^{-1/2} \simeq 2.4 \times10^{18}\,
\mathrm{GeV}$.
The signature of the metric is followed~$(-,+,+,+)$ and the background is spatially flat
Friedmann-Lema\^itre-Robertson-Walker~(FLRW).
An overdot and an overprime shall denote differentiation with 
respect to the cosmic time~$t$ and the conformal time~$\eta$, respectively. Derivatives with respect to any other quantity will be denoted as subscript by the quantity.


\section{Inflation and primordial scalar perturbations}\label{sec:NBD-state}

Let us start with the discussion of inflation driven by a single scalar field minimally coupled to gravity. If the field is described by a potential $V(\phi)$, the equation of motion will be given by 
\begin{align}
    \ddot\phi+3H\dot\phi+V_{\phi}=0,
\end{align}
where $H=\dot{a}/a$ is the Hubble parameter with $a$ being the scale factor. If the velocity of the field remains small, the acceleration term becomes negligible and one obtains the attractor behaviour of the field, $3H\dot\phi=-V_{\phi}$. The slow roll 
parameters can be defined in the following manner
\begin{align}
    \epsilon=\f{\Mpl^2}{2}\l(\f{V_{\phi}}{V}\r)^2,\quad\eta=\Mpl^2\f{V_{\phi\phi}}{V}
\end{align}
which quantify the dynamics of the field. Note that during inflation $\epsilon \ll1$ and the end of inflation typically refer to when $\epsilon$ reaches unity.

One key aspect of inflation is its ability to amplify quantum vacuum fluctuations to classical perturbations. These classical perturbations serve as the seeds for various observable features in the present universe, such as the anisotropies in the CMB and the formation of large-scale structures.
Let $f_k$ be the Fourier mode of the scalar perturbation. The equation of motion of the Mukhanov-Sasaki variable associated with the mode, $v_k=z\,f_k$ with $z=a\Mpl\sqrt{2\epsilon}$ is given by ~\cite{Sasaki:1986hm,Mukhanov:1987pv}
\begin{align}
    v_k''+ \l(k^2-\f{z''}{z}\r)v_k=0.
\end{align}
The power spectrum is defined as the two-point correlation of the perturbative modes. In a slow roll inflation, the power spectrum has the form 
\begin{align}
    \ps(k)=\f{k^3}{2\pi^2}\f{\vert v_k \vert^2}{z^2} = \As\l(\f{k}{k_\ast}\r)^{\ns-1},
\end{align}
where $k_\ast$ is the pivot scale, $\As\simeq 2.09\times10^{-9}$ is the amplitude of the spectrum and $\ns$ is the spectral tilt as mentioned before. For perturbations generated in BD vacuum, $\ns$ and $r$ can be determined using the slow roll parameters as 
\begin{align}
    \ns=1-6\epsilon+2\eta, \quad r=16\epsilon.
    \label{eq:ns-r}
\end{align}
Further, under slow roll approximation, the energy scale of inflation can be related to the tensor-to-scalar ratio $r$ as $\HI=\pi\Mpl\sqrt{r\,\As/2}$.  

Let us now incorporate the effects arising from a NBD initial state. The solution for the Mukhanov–Sasaki variable in the presence of a general NBD initial state can be expressed as:~\cite{Brandenberger:2002hs,Sriramkumar:2004pj,Holman:2007na,Meerburg:2009ys,Agullo:2010ws,Kundu:2011sg}
\begin{align}
    v_k=\alpha_k\,v_k^{\rm BD}+\beta_k^\ast\,v_k^{\rm BD\,\dagger},
\end{align}
where $v_k^{\rm BD}$ denotes the solution corresponding to the standard BD initial condition. Naturally, the choice $\alpha_k=1$ and $\beta_k=0$ recovers the BD case.
Furthermore, the Wronskian condition imposes the constraint:
\begin{align}
    \vert \ak\vert^2-\vert\bk\vert^2=1,
    \label{eq:wrons-alb}
\end{align}
which must be satisfied for the mode functions to preserve the canonical commutation relations.
Under the effect of the NBD initial condition, the scalar power spectrum modifies to 
\begin{align}
    \ps^{\rm NBD}(k)=\ps(k)\,\gk
\end{align}
where the quantity $\gk=1+2\bk^2-2\bk\sqrt{1+\bk^2}$ is the modification due to the NBD case. Here, the condition in Eq.~\eqref{eq:wrons-alb} is used to express $\gk$ completely in terms of $\bk$. Note that we have considered the angle between $\ak$ and $\bk$ to be $\pi$. Thus, under the effects due to NBD initial state, $\ns$ and $r$ are modified to
\begin{align}
    \ns^{\rm NBD}-1&=\ns-1+\f{2}{\gk}
    \f{2\bk\sqrt{1+\bk^2}-(1+\bk^2)}{\sqrt{1+\bk^2}}\f{\d\bk}{\d\ln k},
\end{align}
and
\begin{align}
    \rnbd=\f{r}{\gk}.
\end{align}
The parameter $\bk$ needs to be such that after a certain cutoff scale, the mode restores the BD state. 
A choice of physical model, which is often used for $\bk$ is given by~\cite{Holman:2007na,Ashoorioon:2013eia,Ashoorioon:2014nta,Brahma:2025dio}
\begin{align}
    \bk=\beta_0\e^{-k^2/(a^2M^2)},
    \label{eq:betak}
\end{align}
where $M$ is the physical cutoff, which during inflation, needs to be 
larger than the Hubble parameter. Furthermore, the choice of $\bk$
provides $\d\bk/\d\ln k=-2\bk k^2/(a^2M^2)$~\cite{Ashoorioon:2013eia,Ashoorioon:2014nta,Brahma:2025dio}.
At this point, it is useful to note that, to stay in the linear regime and to avoid backreaction, $\bk\ll1$ is required~\cite{Brahma:2025dio}. 
In~\cite{Porrati:2004gz,Greene:2004np,Ashoorioon:2017toq}, the author demonstrated that the back reaction is not very constraining for corrections to the power spectrum
as long as the scale is different from the Planck scale. The upper value on $\beta_k$ is estimated to be  $H\Mpl\sqrt{\epsilon\eta'}/M^2$. Hence, a 
large value of $\beta_0$ can be obtained by changing the value of $H/M$.
{ In this work, we shall stick to the choice of the cutoff at pivot scale to be $aM\simeq5k_\ast$ to ensure that the pivot scale is well inside the Hubble radius at the initial time~\cite{Brahma:2025dio}. Further,  we shall limit choosing $\beta_0<0.2$ to avoid backreaction (a discussion regarding the choice of the value is given in Sec.~\ref{sec:lbk})}.
In the next section, we shall discuss the effects due to the phase of reheating on the inflationary observables.


\section{Constraints from the phase of reheating}\label{sec:reheating}

As already mentioned, the phase of reheating connects the inflation to the epoch of standard radiation domination.
It is a phase with the energy density dominated by inflation, and during which the field transfers its energy through various couplings to the SM particles. 
There exist several processes through which the inflaton can transfer its energy to daughter particles. This transfer can occur either via non-gravitational couplings~\cite{Garcia:2020wiy}, or through purely gravitational interactions~\cite{Clery:2021bwz,Clery:2022wib,Haque:2022kez,deHaro:2024gml} (where no additional coupling is required and reheating occurs purely due to time-dependent and curved background). In addition, ultralight primordial black holes may form during reheating, which can subsequently emit SM particles through Hawking evaporation~\cite{RiajulHaque:2023cqe,Maity:2024cpq}. In this work, we focus on studying the reheating phase in a model-independent manner, without delving into the specifics of the couplings.
In general, the phase of reheating can be characterized by the following variables: reheating EoS, $\wre$, reheating temperature, $\Tre$, and the duration of reheating, $\Nre$.
The end of reheating is defined when $\rho_{\phi}=\rho_{_{\rm R}}=\rhore$, where $\rho_{\phi}$ and $\rho_{_{\rm R}}$ are the energy density of inflation and radiation respectively. Hence, one can use the 
conservation of entropy, $g_{\rm s}a^3T^3=\text{constant}$
at the end of reheating, where $g_{\rm s}$ is the relativistic degrees of freedom associated with the entropy. 
Using the conservation relation, we get~\cite{Dai:2014jja,Cook:2015vqa}
\begin{align}
    \Tre=\l(\f{43}{11\,g_{\rm s}^{\rm re}}\r)^{1/3}\f{\HI}{k_\ast}\e^{-(\Nk+\Nre)},\label{eq:Tre1}
\end{align}
where we have used, at present day $g_{\rm s}^0=43/11$. The quantity $\Nk$ is the duration from the pivot scale to the end of inflation, and $\HI$ is the Hubble parameter at the end of inflation.
Further, in a background driven by a general EoS, $\wre$, the total energy density falls as $\rho\propto a^{-3(1+\wre)}$.
Hence, using the relation $\rhore=(\pi^2/30)g_{\rm re}\Tre^4$,
the temperature of reheating is given by 
\begin{align}
    \Tre=\l(\f{90\Mpl^2\HI^2}{\pi^2g_{\rm re}}\r)^{1/4} \e^{-(3/4)(1+\wre)\Nre}.
    \label{eq:Tre2}
\end{align}
It is straightforward to see from Eq.~\eqref{eq:Tre2}, that $\Nre$, $\wre$ and $\Tre$ are not completely independent. One can be obtained if the other two quantities are specified.
Combining Eqs.~\eqref{eq:Tre1} and \eqref{eq:Tre2} we get the duration of inflation in terms of $\wre$ and $\Tre$ to be
\begin{align}
    \Nk&= \log\l[\l(\f{43}{11\,g_{\rm s}^{\rm re}}\r)^{1/3}
    \l(\f{\pi^2\,g_{\rm re}}{90\Mpl^2\HI^2}\r)^{\f{1}{3(1+\wre)}}\r.\nn\\
    & \l.\quad\quad\quad
    \times T_0\f{\HI}{k_\star} \Tre^{\f{1-3\wre}{3(1+\wre)}}
\r],\label{eq:Nk}
\end{align}
where $T_0$ is the temperature of the CMB at the present day. 
From Eq.~\eqref{eq:Nk}, is straightforward to see that, for $\wre=1/3$, $\Nk$ will be independent of $\Tre$. 
For $\wre<1/3$, a higher reheating temperature will lead to a higher value of $\Nk$, whereas the dependence becomes opposite for $\wre>1/3$.
In slow roll inflation, $\Nk$ becomes extremely important to determine the value of $\phi$ at the pivot scale. Which further is used to determine $\epsilon$, $\eta$, and $\ns$. The details of obtaining $\Nk$ shall be discussed later. In the next section, we shall give a brief overview of the bound of $\dneff$ to the strength of the GW spectrum. 


\section{Constraints from the primary GWs and $\dneff$}\label{sec:pgw-dneff}

The spectral energy density of primary GWs, which is defined as the ratio between the energy density of GWs to the critical energy density of the universe, $\ogw=\rhogw / \rho_{\rm c}$ is given by~\cite{Mishra:2021wkm,Haque:2021dha,Maity:2024cpq}  
\begin{align}
    \ogw(k)h^2& =c_g\,\oR h^2\f{\HI^2}{12\pi^2\Mpl^2}
    \f{4\mu^2}{\pi} \nn\\&\quad\times
    \Gamma^2\l(1+\f{\nu}{\mu}\r) \l(\f{k}{2\mu\kre}\r)^{n_{_\mathrm{GW}}},
    \label{eq: gw energy density final}
\end{align}
where the quantities $\mu=(1/2)(1+3\wre)$ and $\nu=(3/4)(1-\wre)$.
The quantity $c_g=(g_{\rm eq}/g_{0})(g_{\rm s}^{\rm eq}/g_{\rm s}^{0})^{4/3}$ contains the combination of relativistic degrees of freedom. Also, $\oR$ is the energy density of the radiation today. The spectral index of the GWs is given by $n_{_\mathrm{GW}}=-2(1-3\wre)/(1+3\wre)$. It is clear that during radiation dominated epoch ($\wre=1/3$), we obtain a scale-invariant spectrum. We also see that reheating with  $\wre<1/3$ will induce a red tilt and $\wre>1/3$ will induce a blue tilt to the spectrum for the modes that enter during the phase of reheating, i.e., for $\ke>k>\kre$. 
{ At this stage, we clarify that the exact form of the tensor power spectrum can be obtained by considering the evolution of a specific model~\cite{Kinney:2021nje,Giare:2022wxq}.
Here we have considered the de Sitter approximation, under which inflation is assumed to generate a scale-invariant tensor power spectrum throughout.}

Our next goal is to check after fixing the reheating EoS and the energy scale of the inflation, how small $\Tre$ can reach in order to be consistent with the $\dneff$ constraints.
As discussed, $\dneff$ is the additional relativistic degrees of freedom at the time of BBN or CMB decoupling. In the case of the GWs the expression of $\dneff$ can be written as \cite{Jinno:2012xb}
\begin{align}
    \dneff=\frac{\rhogw}{\,\rho_{\nu}}
   =\frac{8}{7}\left(\frac{11}{4}\right)^{\frac{4}{3}}\frac{\rhogw}{\rho_{\rm \gamma}}\,,
   \label{Eq: neff}
\end{align}
where $\rho_\nu$ and $\rho_{\rm \gamma}$ are the energy density of single SM neutrino species and the energy density of photon, respectively. 
We have used the relation between the temperature of neutrino and photon, $T_{\rm \nu} = \left({4}/{11}\right)^{{1}/{3}}\, T_{\rm \gamma}$ in Eq.~\eqref{Eq: neff}. The above equation provides the restriction on the current energy density of GWs as\cite{Chakraborty:2023ocr}
\begin{align}
  \int_{\kre}^{k_{\rm f}}\frac{dk}{k}\Omega_{_{\rm GW}}h^2(k)\leq \frac{7}{8}\left(\frac{4}{11}\right)^{4/3}\Omega_{\rm \gamma}h^2\,\dneff,
  \label{eq: deltaneff}
\end{align}
where $\Omega_{\rm \gamma}h^2\simeq 2.47\times10^{-5}$ is the photon relic density at the present day.
It is straightforward to notice that the bound from $\dneff$ becomes effective only for a blue-tilted spectrum which is obtained for a steeper EoS, i.e. $\wre>1/3$.
In such a case, upon substitution of Eq.~\eqref{eq: gw energy density final}  in Eq.~\eqref{eq: deltaneff} and performing the integration, one can find
\begin{align}
  \int_{\kre}^{\ke}\frac{\d k}{k}\ogw (k)h^2 &\simeq c_g\,\oR h^2\f{\HI^2}{12\pi^2\Mpl^2} \Gamma^2\l(\f{5+3\wre}{2+6\wre}\r)
  \nn\\&\quad\times
  \f{(1+3\wre)^{1+4/(1+3\wre)}}{2\pi(3\wre-1)}\l(\f{\ke}{\kre}\r)^{\f{6\wre-2}{1+3\wre}},\nn\\
  &\simeq c_g\,\oR h^2\f{\HI^2}{12\pi^2\Mpl^2}\zeta(\wre)\l(\f{\ke}{\kre}\r)^{\f{6\wre-2}{1+3\wre}}
  \label{eq: BBNapprox}
\end{align}
in Eq.~\eqref{eq: BBNapprox}, the ratio between $\ke$ and $\kre$ can be expressed in terms of $\HI$, $\wre$ and $\Tre$ as
\begin{align}
\frac{\ke}{\kre}=\left(\frac{90}{\pi^2 g_{\rm re}}\f{\Mpl^2\HI^2}{\Tre^4}\right)^{\frac{1+3\, \wre  }{6\,(1+ \wre  )}}.
\label{eq: kend kre}
\end{align}
Combining Eq.~\eqref{eq: deltaneff}, Eq.~\eqref{eq: BBNapprox} and Eq.~\eqref{eq: kend kre}, we obtain the restriction on the lower bound of the reheating temperature as
\begin{align}
\Tre &\geq \left(\frac{90\Mpl^2\HI^2}{\pi^2 g_{\rm s}^{\rm re}}\right)^{1/4}
\nn\\&\quad\times\left(\frac{c_g \oR h^2\, \zeta( \wre  )}{5.61\times 10^{-6}\,\Delta N_{\rm eff}}\f{\HI^2}{12\pi^2\Mpl^2}\right)^{\frac{3\,(1+ \wre  )}{4\,(3\, \wre  -1)}} .
\label{eq:BBNrestriction}
\end{align}
In Eq.~\eqref{eq:BBNrestriction}, $\dneff$ is tightly constrained by recent ACT data, yielding an upper limit of $\dneff \leq 0.17$ at $95\%$ confidence level~\cite{ACT:2025fju,ACT:2025tim}. It is important to note that there also exists a lower bound on the reheating temperature, set by BBN, which requires $\Tre \gtrsim 4$ MeV~\cite{Kawasaki:1999na,Kawasaki:2000en,Hasegawa:2019jsa}. It can be shown that the bound on $\Tre$ derived from Eq.~\eqref{eq:BBNrestriction} is stronger than the BBN bound only for equations of state with $\wre > 0.58$. Therefore, in our analysis, we consider the minimum reheating temperature to be max[$\Tre,T_{_\mathrm{BBN}}$], while the maximum reheating temperature corresponds to the case of instantaneous reheating, which can be directly computed from the energy scale at the end of inflation. With these tools in hand, we now turn to a discussion of the slow roll models of interest in the next section.


\section{Slow roll models}\label{sec:models}

In this section, we consider a selection of single-field slow roll inflationary models that are widely studied in the literature due to their theoretical motivation and compatibility with observational data. This set includes plateau-like potentials such as the $\alpha$-attractor E-model and T-model. We also examine chaotic inflation models characterized by monomial potentials of the form $\phi^n$, along with other well-motivated scenarios like hilltop inflation, where inflation occurs near the top of a potential hill, and natural inflation. These models are well motivated and provide a representative sample for exploring the impact of recent cosmological constraints on the dynamics of inflation.

Under the slow roll approximation, the duration of inflation, $\Nk$ can be computed using the following relation:
\begin{align}
    \Nk=-\f{1}{\Mpl^2}\int_{\phi_k}^{\phie}
    \d\phi \f{V}{V_\phi},\label{eq:nki}
\end{align}
where $\phie$ is the value of the field at the end of inflation, can then be determined by equating the expression for  $\epsilon
(\phie)=1$. On the other hand, $\phi_k$ can be calculated by comparing $\Nk$  from Eq.~\eqref{eq:Nk} and Eq.~\eqref{eq:nki}. Further, 
$\ns$ and $r$ can be calculated  using Eq.~\eqref{eq:ns-r} with the substitution of $\phi_k$.
Lastly, the overall constant of the potential, $V_0$ can be obtained by using the following relation
\begin{align}
    V(\phi_k)=\f{3}{2}\pi^2\Mpl^4r\As.
    \label{eq:vphik}
\end{align}
Our approach in this paper is as follows. We examine the inflationary models listed previously, considering different reheating scenarios characterised by an effective EoS. For each model, we analyse whether it can satisfy the $\ns-r$ constraints from the combined P-ACT-LB-BK18 dataset over the allowed range of reheating temperatures $\Tre$. Naturally, a significant portion of the parameter space, defined by both the parameters of inflation and reheating, will remain excluded even after introducing the phase of reheating.
To further explore the viability of the models, we introduce NBD initial conditions and examine whether the resulting corrections can bring $\ns$  into agreement with observational bounds. It is important to note, however, that modifying the initial vacuum state also alters the tensor-to-scalar ratio $r$. Therefore, we must simultaneously ensure that any improvement in $\ns$
  through the NBD initial state does not lead to a violation of the upper bound on $r$.

\begin{figure*}
    \centering
    \includegraphics[width=0.47\linewidth]{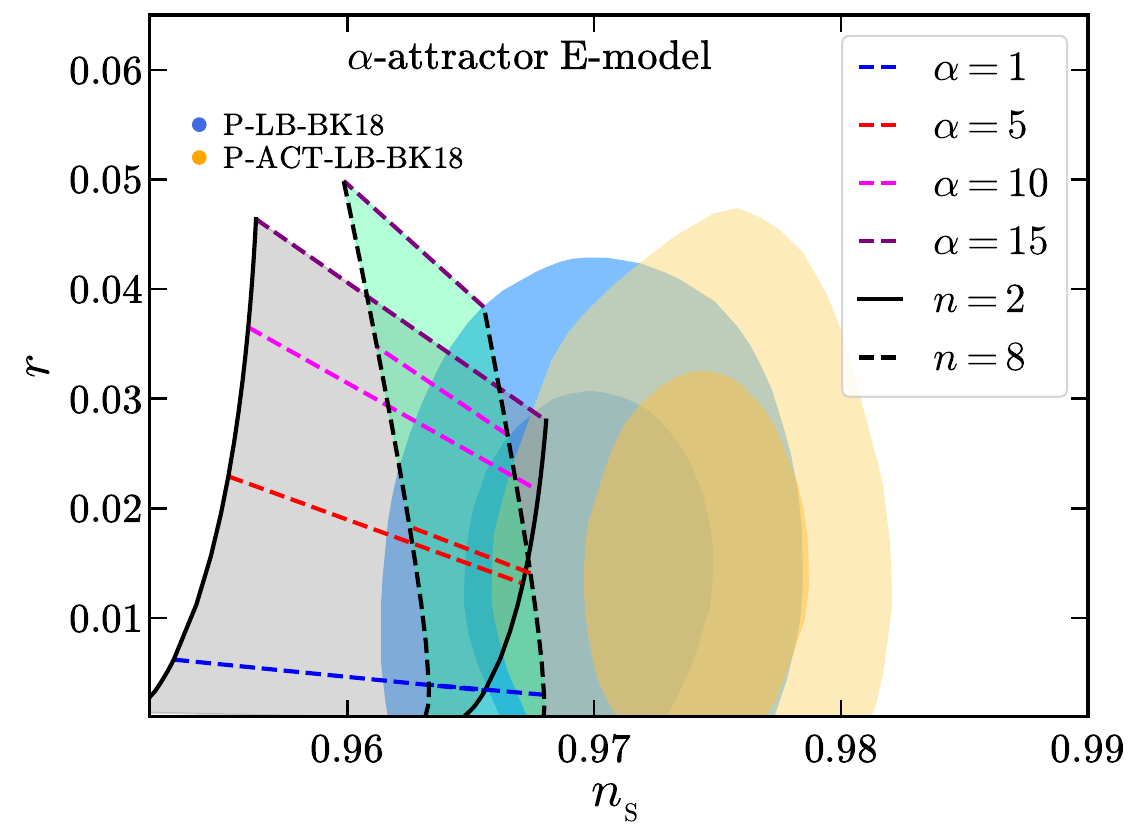}
    \includegraphics[width=0.47\linewidth]{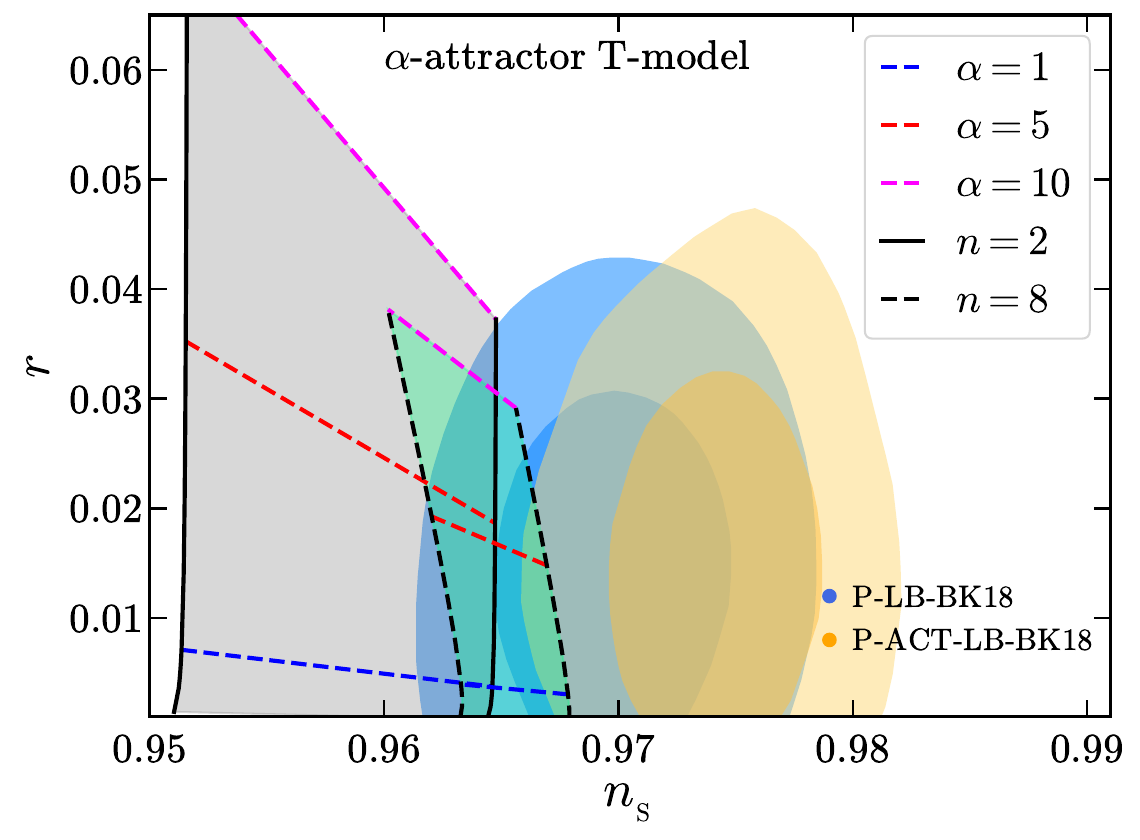}
    \includegraphics[width=0.47\linewidth]{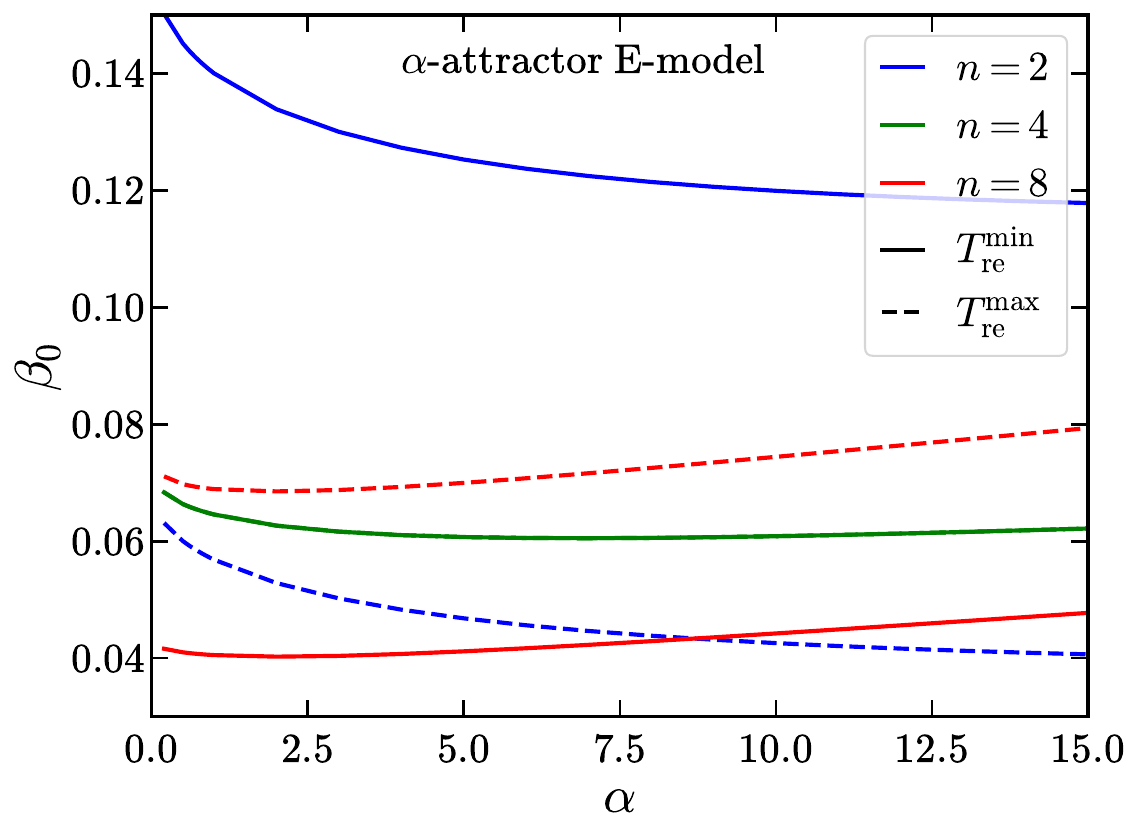}
    \includegraphics[width=0.47\linewidth]{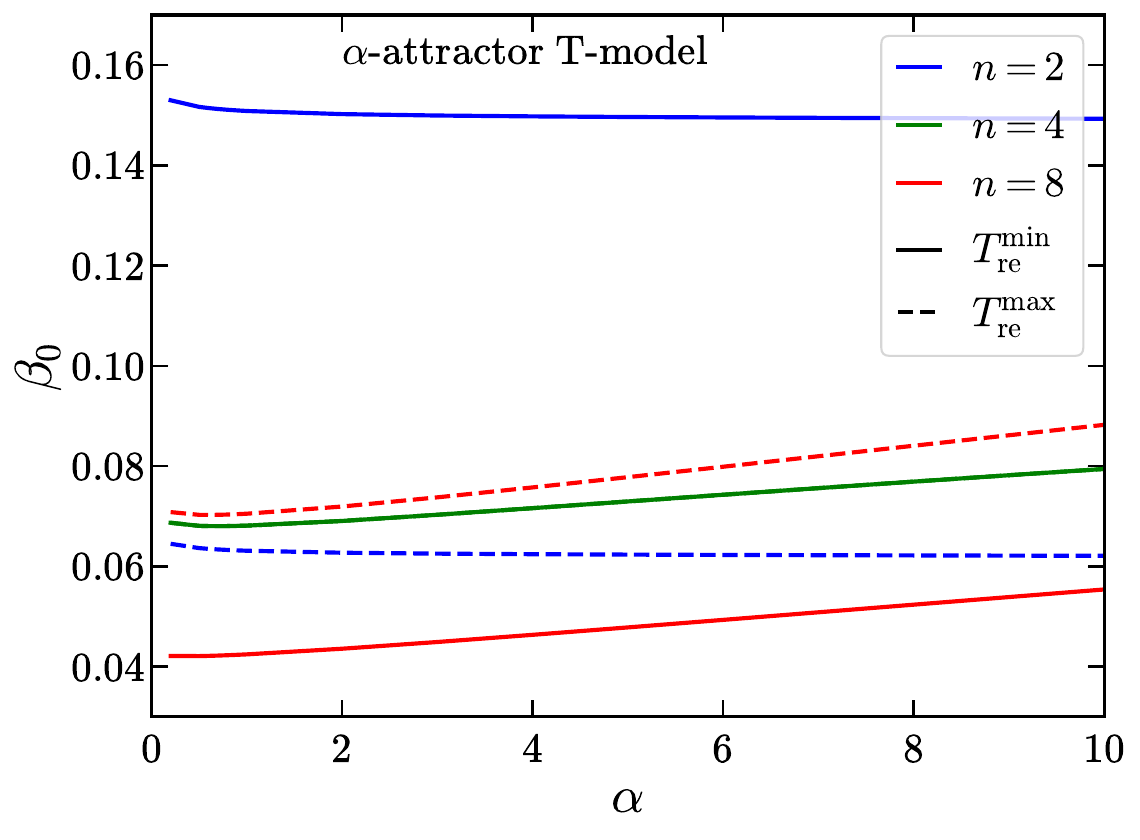}
    \includegraphics[width=0.47\linewidth]{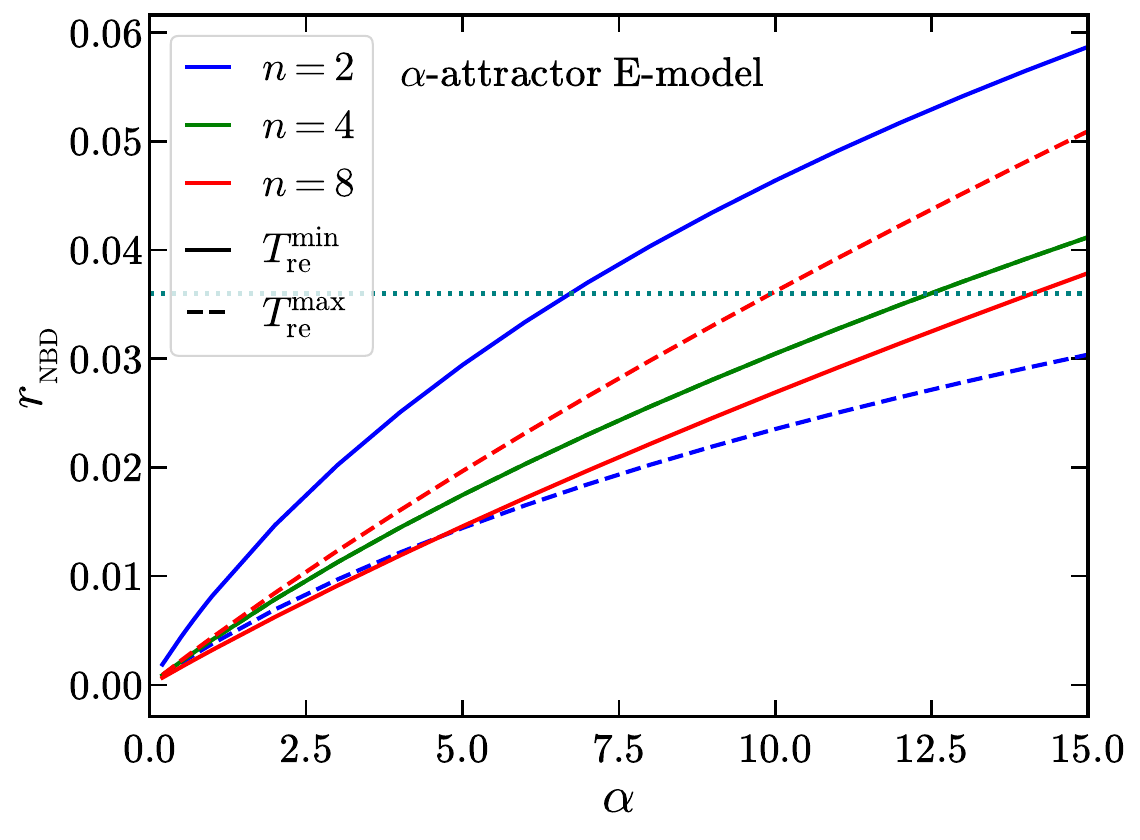}
    \includegraphics[width=0.47\linewidth]{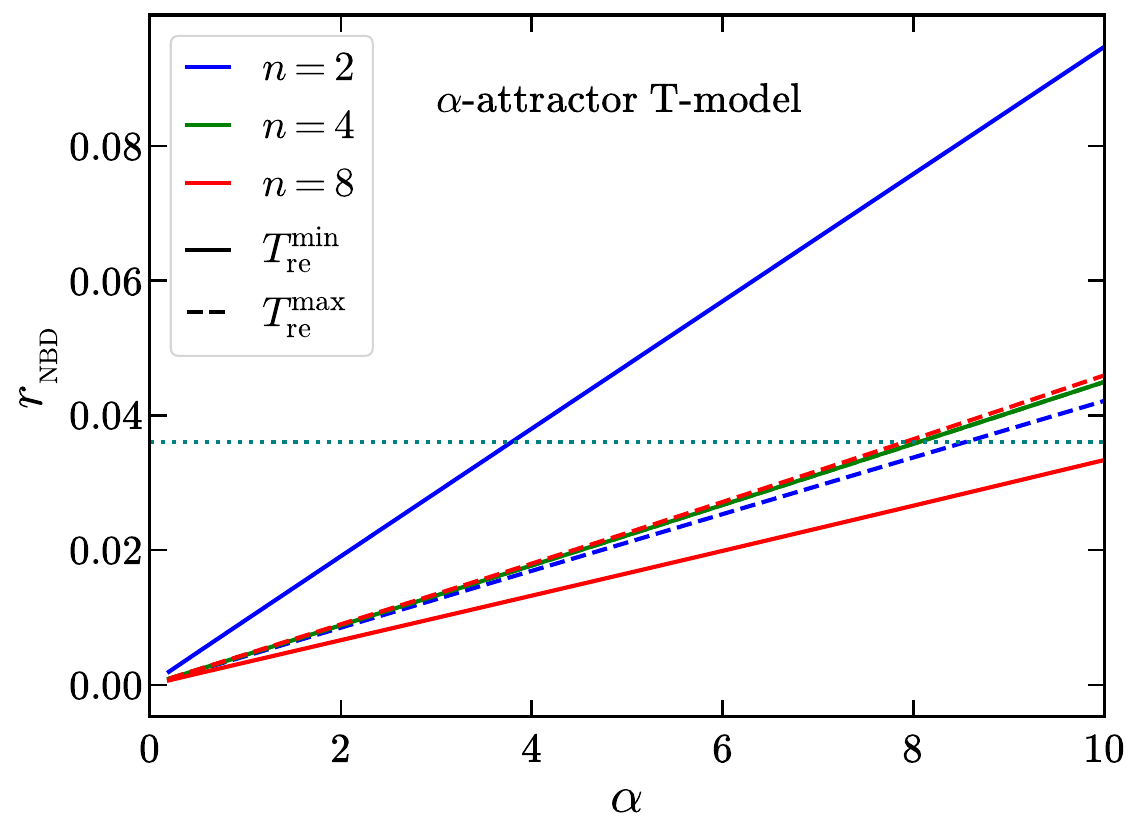}
    \caption{The prediction of $\ns-r$ for $\alpha$-attractor E-model (top left) and $\alpha$-attractor T-model (top right) are plotted along with the constraints provided by P-LB-BK18 (blue) and P-ACT-LB-BK18 (yellow).
     The darker and light shaded areas in the contour indicate $1\sigma$ region at $68\%$ C.L. and $2\sigma$ region at $95\%$ C.L. 
     { Note that the constraints are drawn using the data of~\cite{ACT:2025fju} as also given in~\cite{Kallosh:2025rni}.}
    The dashed blue, red, magenta and purple lines correspond to $\alpha=1$, $5$, $10$ and $15$ respectively.
    The two bounds in black correspond to the highest and lowest allowed $\Tre$. The gray and green shaded areas correspond to $n=2$ and $n=8$.
    The middle plots correspond to the value of $\beta_0$ for $\alpha$-attractor E-model (middle left) and for $\alpha$-attractor T-model (middle right), that is required in order to satisfy the mean value of $\ns=0.9743$ as provided by the current ACT data.  
    In the bottom plots, the tensor-to-scalar ratios are plotted which are obtained after incorporating the modification from the NBD initial state in order to satisfy $\ns=0.9743$. 
    }
    \label{fig:a-model}
\end{figure*}



\subsection{$\alpha$-attractor models}

Let us start by considering a specific model of inflation that permits slow roll evolution. The $\alpha$-attractor E-model is a class of inflationary potentials derived from supergravity frameworks with hyperbolic field space geometry.
The potential $V(\phi)$ takes the form: \cite{Starobinsky:1980te,Starobinsky:1983zz,Kallosh:2013hoa,Kallosh:2013yoa}
    \begin{align} \label{a}
V(\phi) = V_0 \left[  1 - \exp\l({ -\sqrt{\f{2}{3\alpha}}\frac{\phi}{\Mpl} }\r) \right]^{n}\,,
\end{align}
where $\alpha$ determines the curvature of the field space and $n$ is a positive integer.
For large field values, the potential becomes asymptotically flat, producing a plateau that leads to prolonged inflation.
It is also straightforward to see that for $\alpha=1$ and $n=2$ the predictions from this potential approach those of the Starobinsky model.

The $\alpha$-attractor T-model is another realization of $\alpha$-attractors where the potential is symmetric around the origin and given by~\cite{German:2021tqs}
\begin{align}
    V(\phi)=V_0 \tanh^n\l(\f{1}{\sqrt{6\alpha}}\f{\phi}{\Mpl}\r),
\end{align}
where, similar to E-model, $\alpha$ determines the curvature of the field space and $n$ is a positive integer.
For more references related to $\alpha$-attractors, see~\cite{Odintsov:2016vzz,Ueno:2016dim,Kumar:2015mfa,Eshaghi:2016kne,Dalianis:2018frf}.
In both models, $V_0$ denotes the inflationary energy scale which can be determined using the relation in Eq.~\eqref{eq:vphik}. 
The slow roll parameters in the E-model are given by 
\begin{align}
     \epsilon &=\f{n^2}{3\alpha}\l[\exp\l(\sqrt{\f{2}{3\alpha}}\f{\phi}{\Mpl}\r)-1\r]^{-2}\\
     \eta &= \f{2n}{3\alpha}
     \l[n-\exp\l(\sqrt{\f{2}{3\alpha}}\f{\phi}{\Mpl}\r)\r]\nn\\&\quad\times
     \l[\exp\l(\sqrt{\f{2}{3\alpha}}\f{\phi}{\Mpl}\r)-1\r]^{-2}
 \end{align}
 and for the T-model, the parameters have the following form 
\begin{align}
    \epsilon&=\f{n^2}{3\alpha} \mathrm{cosech}^2\l(\sqrt{\f{2}{3\alpha}}\f{\phi}{\Mpl}\r)\\
    \eta &= \f{n}{6\alpha}
    \mathrm{sech}^2\l({\f{1}{\sqrt{6\alpha}}}\f{\phi}{\Mpl}\r)
    \nn\\&\quad\times
    \l[(n-1)\f{\mathrm{sech}^2\l({\f{1}{\sqrt{6\alpha}}}\f{\phi}{\Mpl}\r)}{\mathrm{tanh}^2\l({\f{1}{\sqrt{6\alpha}}}\f{\phi}{\Mpl}\r)}-2\r].
\end{align}
The spectral index and the tensor-to-scalar ratio can be found using Eq.~\eqref{eq:ns-r}. Finally the number of e-fold from when the pivot scale leave the Hubble radius to the end of inflation is given in E-model as 
\begin{align} 
N_k &= \f{3\alpha}{4n} \l[\exp\l(\sqrt{\f{2}{3\alpha}} \f{\phi_k}{\Mpl}\r)  - \exp\l(\sqrt{\f{2}{3\alpha}}\f{\phi_{\rm end}}{\Mpl}\r)\r. \nn\\& \l. \quad\quad
-\sqrt{\f{2}{3\alpha}} \f{\phi_k-\phi_{\rm end}}{\Mpl}\r]\,,
\label{eq:Nk-e}
\end{align} 
and in T-model as 
\begin{align}
    N_k=\f{3\alpha}{2n}\l[\cosh\l(\sqrt{\f{2}{3\alpha}}\f{\phi_k}{\Mpl}\r)-\cosh\l(\sqrt{\f{2}{3\alpha}}\f{\phie}{\Mpl}\r)\r].
    \label{eq:Nk-t}
\end{align}
A recent study has demonstrated that, in light of the P-ACT-LB-BK18 data, the Starobinsky model of inflation~\cite{Starobinsky:1980te} is pushed toward the edge of the allowed $\ns-r$ region, lying close to the $2\sigma$ contour~\cite{Kallosh:2025rni}. We begin by investigating whether this tension can be alleviated through a subsequent phase of reheating. At this point, it is important to highlight a key consideration: in such analyses, the reheating EoS parameter can often be treated as a free parameter, 
{ without directly linking $\wre=(n-2)/(n+2)$ for a potential behaving $\phi^n$ at minima. In such cases it is considered that after the end of inflation, the potential has different behavior at the minima.}
We will adopt this approach for most of the models considered in this work. However, in the case of the $\alpha$-attractor models, for simplicity and to reduce the number of free parameters, we shall assume that the reheating EoS is determined by the behavior of the inflaton field near the minimum of the potential. Specifically, if the potential near the minimum behaves as $V(\phi)\sim \phi^n$, then the effective EoS is given by $\wre=(n-2)/(n+2)$~\cite{Garcia:2020wiy}.

In the top panel of Fig.~\ref{fig:a-model}, we show the predictions for $\ns-r$ from both the E- and T-models of the $\alpha$-attractor class, overlaid with the observational contours from P-LB-BK18 and P-ACT-LB-BK18. We consider two representative values of the parameter $n$, which lead to reheating EoS that are, respectively, softer and stiffer than radiation. The allowed regions in the $\ns-r$ plane are enclosed by black curves, which correspond to the constraints from the maximum and minimum reheating temperatures. The upper bound is set by instantaneous reheating, determined by the energy scale at the end of inflation, while the lower bound is given by the BBN temperature, for $n=2$, and for $n=8$, it follows from Eq.~\eqref{eq:BBNrestriction}. We observe that even after accounting for the effects of reheating, the predictions of the $\alpha$-attractor models remain mostly outside the $1\sigma$ region, and thus the reheating phase alone does not significantly improve the model’s agreement with current data.

We now examine the impact of the NBD initial state on the inflationary predictions. Our objective is to achieve the spectral index $\ns=0.9743$, as favored by the P-ACT-LB-BK18 data, by introducing a non-zero value of the Bogoliubov parameter $\beta_0$. In the middle panel of Fig.~\ref{fig:a-model}, we plot the required values of $\beta_0$ as a function of the parameter $\alpha$, ensuring $\ns=0.9743$ is satisfied. We consider three choices of the parameter, where $n=2$ and $n=4$ correspond to reheating equations of state similar to matter and radiation, respectively, while $n=8$ leads to a steeper equation of state.
The solid and dashed curves indicate the bounds from the minimum and maximum allowed reheating temperatures. In the bottom panel of Fig.~\ref{fig:a-model}, we present the corresponding values of the tensor-to-scalar ratio, $r$ as a function of $\alpha$, incorporating the corrections from the NBD initial state.
For $n=2$, we find that a low value of $\Tre$ tends to push $r$ above observational bounds when NBD effects are included. In contrast, higher values of $\Tre$ allow a wide range of $\alpha$ values to remain within the observational constraints. Notably, the Starobinsky model, corresponding to $n=2$ and $\alpha=1$, comfortably satisfies the $\ns-r$ constraints when NBD corrections are taken into account. For $n>4$, the situation is reversed: a lower reheating temperature provides a broader allowed parameter space in $\alpha$ to satisfy the desired $\ns$ while remaining consistent with the bound on $r$ under NBD corrections.

\begin{figure*}
    \centering
    \includegraphics[width=0.47\linewidth]{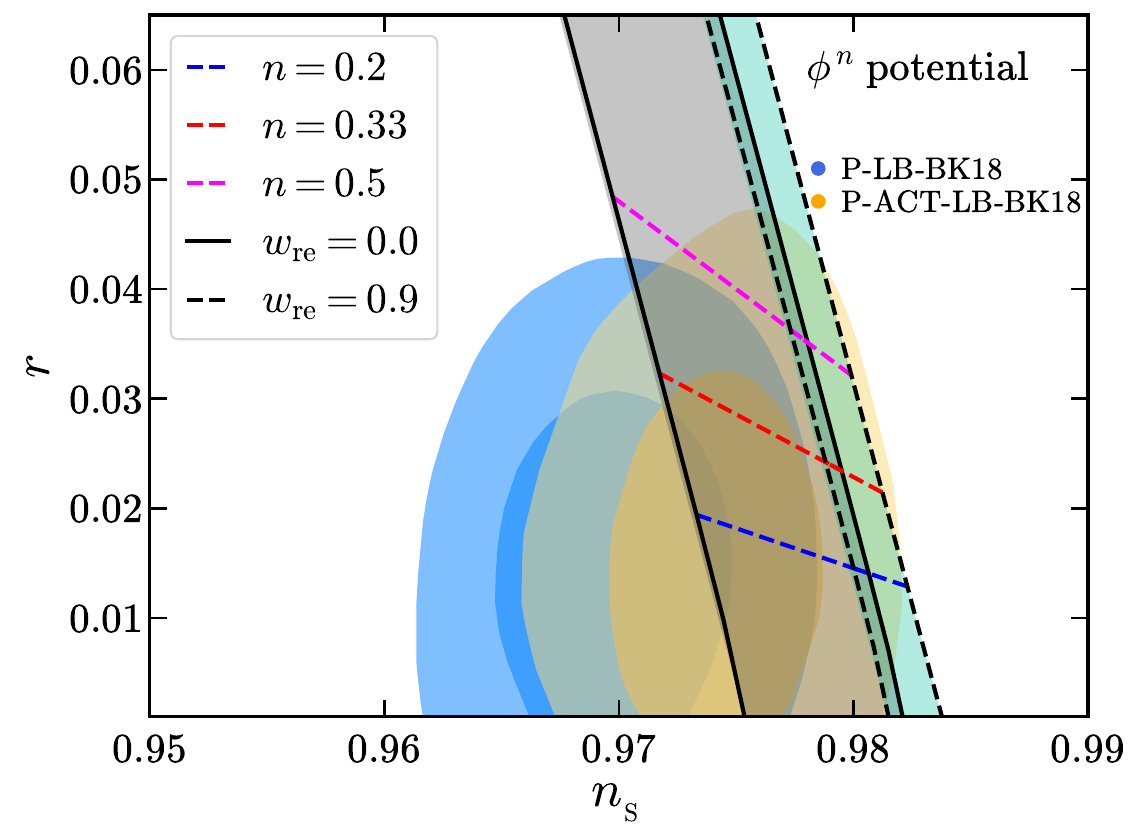}
    \includegraphics[width=0.47\linewidth]{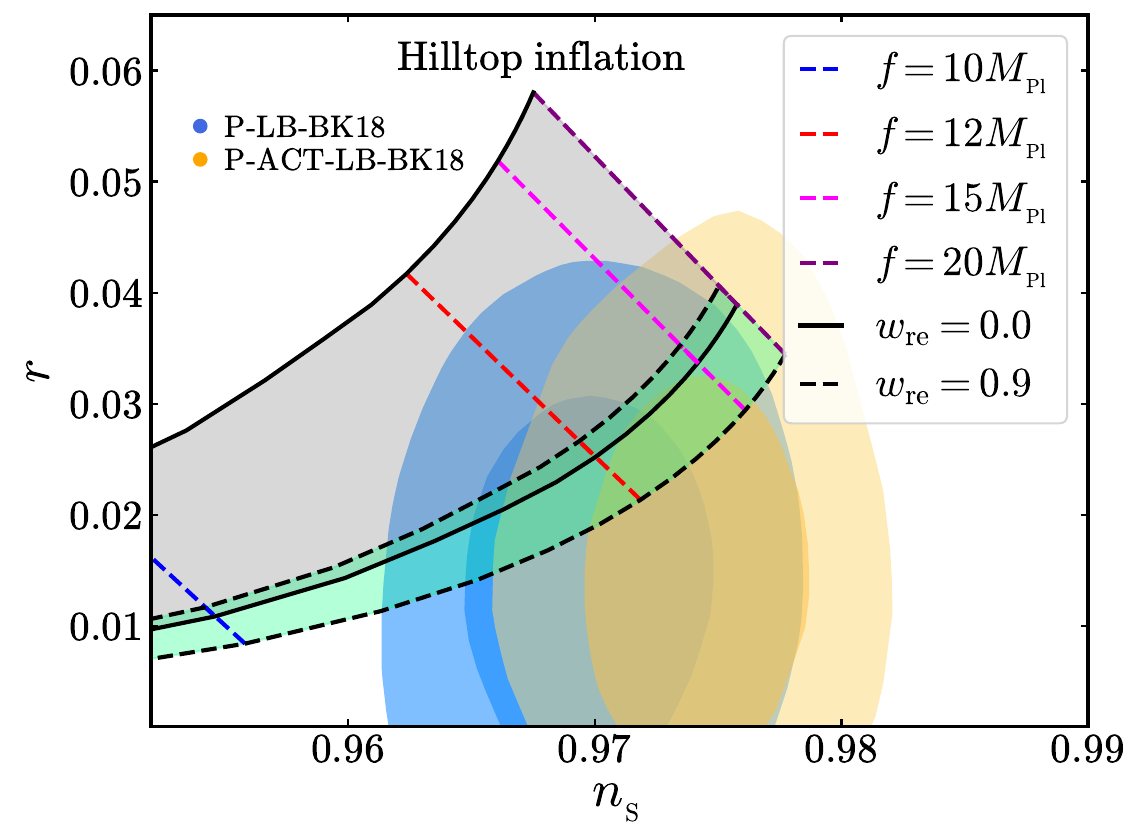}
    \includegraphics[width=0.47\linewidth]{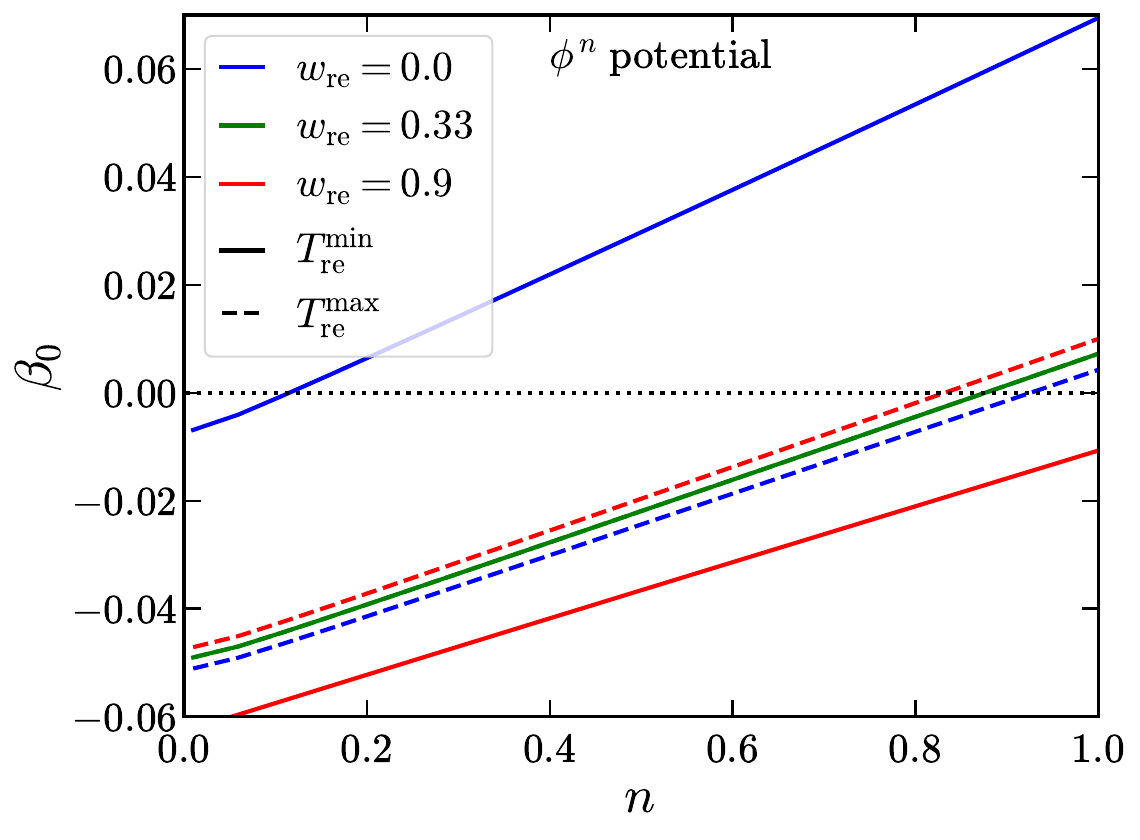}
    \includegraphics[width=0.47\linewidth]{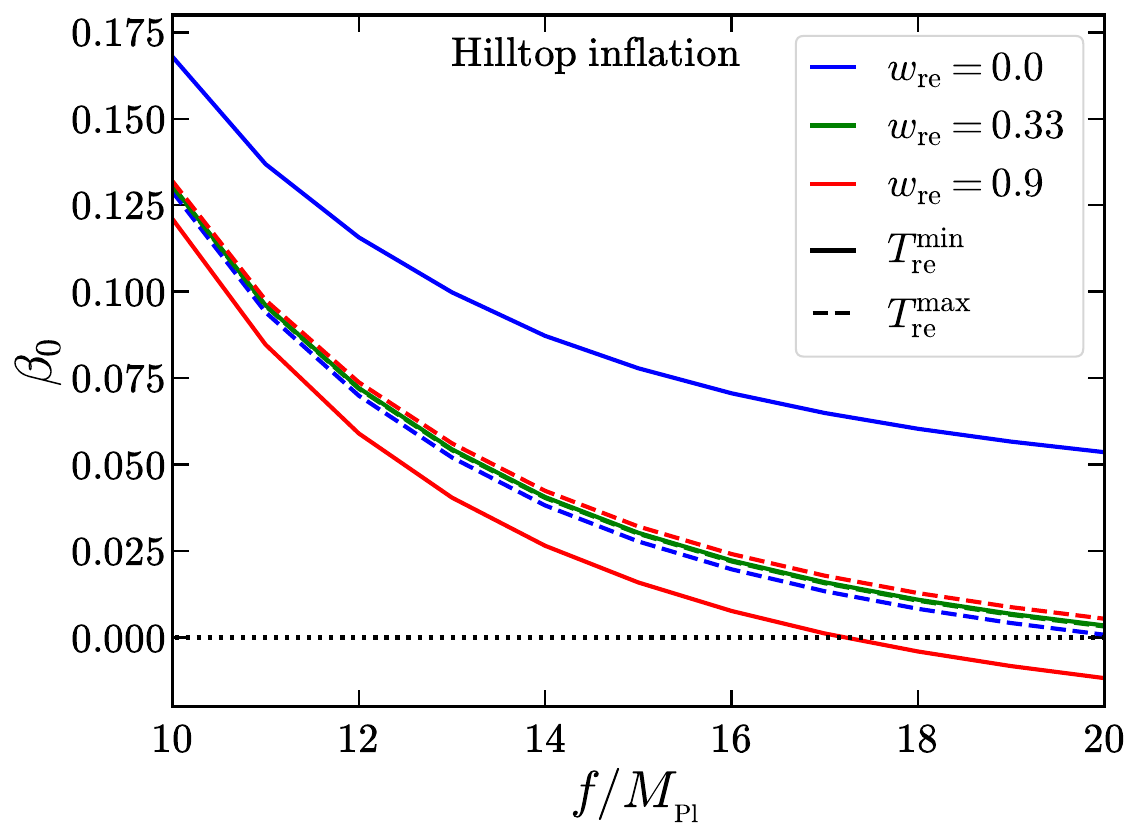}
    \includegraphics[width=0.47\linewidth]{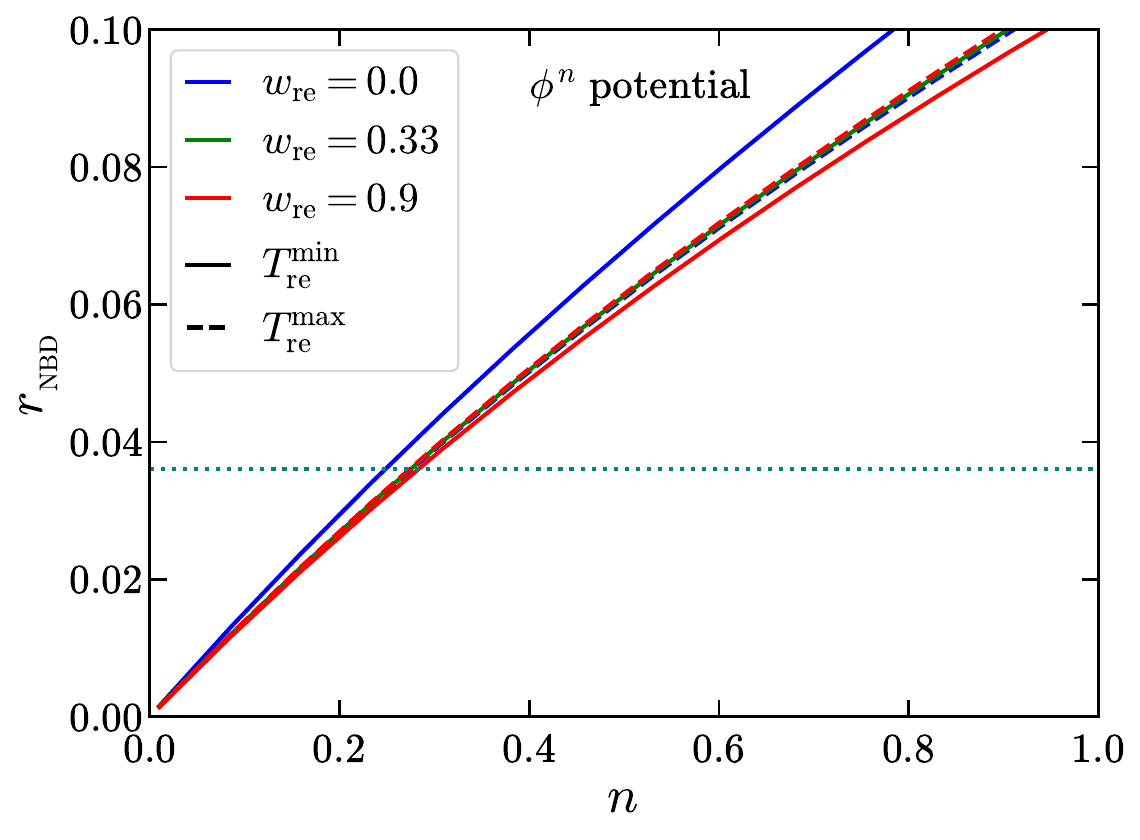}
    \includegraphics[width=0.47\linewidth]{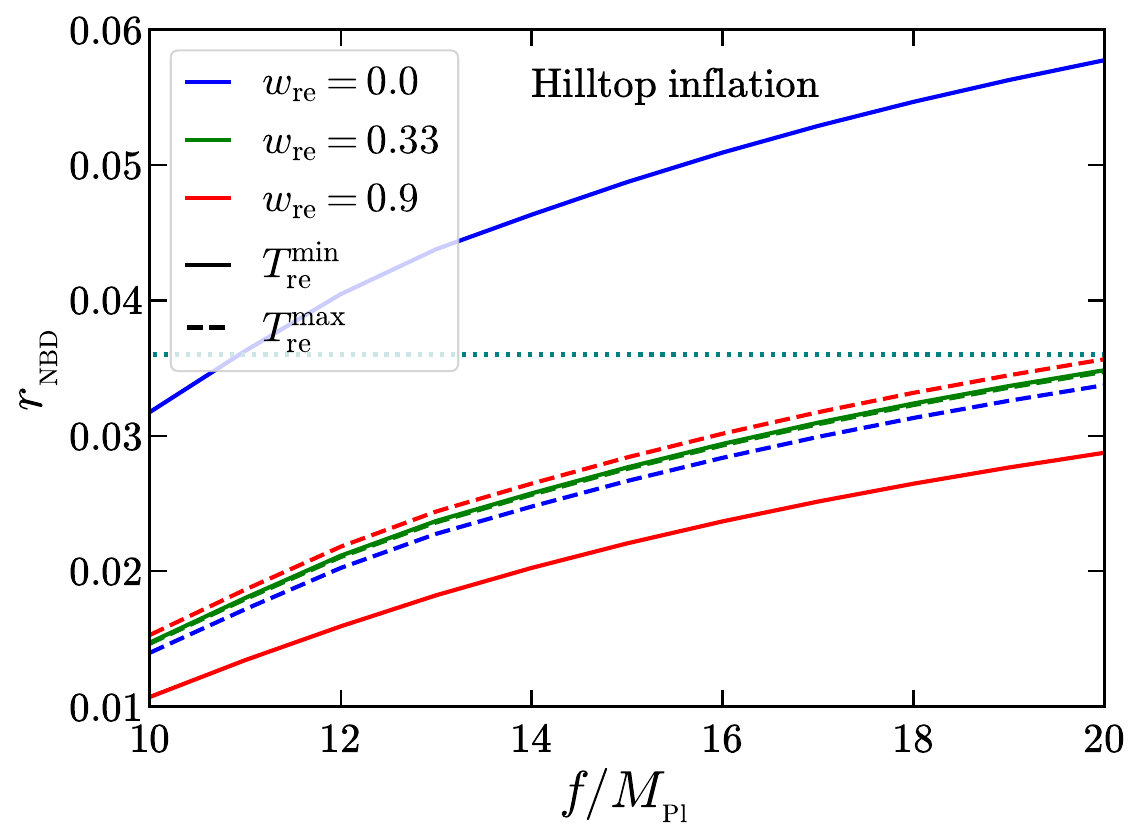}
    \caption{The prediction of $\ns-r$ for $\phi^n$ potential (top left) and hilltop potential (top right) are plotted along with the constraints provided by P-LB-BK18 (blue) and P-ACT-LB-BK18 (yellow).
     The darker and light shaded areas in the contour indicate $1\sigma$ region at $68\%$ C.L. and $2\sigma$ region at $95\%$ C.L.
    The dashed blue, red, magenta lines correspond to $n=1/2$, $1/3$, and $2/3$ in $\phi^n$ potential and the dashed blue, red, magenta, purple lines correspond to $f/\Mpl=10$, $12$, $15$ and $20$ in hilltop model.
    The two bounds in black correspond to the highest and lowest allowed $\Tre$. The gray and green shaded areas correspond to $\wre=0$ and $\wre=0.9$.
    The middle plots correspond to the value of $\beta_0$ for $\phi^n$ potential (middle left) and for hilltop potential (middle right), that is required in order to satisfy the mean value of $\ns=0.9743$ as provided by the current ACT data.  
    In the bottom plots, the tensor-to-scalar ratios are plotted which are obtained after incorporating the modification from the NBD initial state in order to satisfy $\ns=0.9743$.}
    \label{fig:ct-ht}
\end{figure*}
\subsection{$\phi^n$ potential}

Let us now consider the scenarios of chaotic inflation with the potential having the monomial form as~\cite{Linde:1983gd,Martin:2013tda,Cook:2015vqa}
 \begin{align}
    V(\phi)=V_0\l(\f{\phi}{\Mpl}\r)^n,
\end{align}
where $n$ is a positive constant. In our scenario, we shall consider the case of $0.1\leq n \leq2$. In such a model, it is very trivial to obtain the slow roll parameters, which have the following forms
\begin{align}
    \epsilon&=\f{n^2\Mpl^2}{2\phi^2}, \quad
    \eta = \f{n(n-1)\Mpl^2}{\phi^2}.
\end{align}
And the number of e-folds during inflation is given by
\begin{align}
    \Nk=\f{1}{2n\Mpl^2}(\phi_k^2-\phie^2).
\end{align}
In the top left panel of Fig.~\ref{fig:ct-ht}, we show the predictions for the scalar spectral index $\ns$ and tensor-to-scalar ratio $r$ in chaotic inflation models for two representative values of the reheating EoS parameter: $\wre=0$ and $\wre=0.9$. The observational contours from P-LB-BK18 and P-ACT-LB-BK18 are also displayed for comparison. As noted earlier, the black curves delineate the parameter space bounded by the maximum and minimum allowed reheating temperatures, $\Tre$. Previously, in the absence of ACT data, chaotic inflation was nearly ruled out, even for small values of $n$~\cite{Planck:2018jri,Planck:2018vyg,BICEP2:2018kqh,BICEP:2021xfz}. However, the inclusion of ACT, which shifts the central value of $\ns$ upward, has significantly altered the viability of these models. We find that for $n=1/5$ and even $1/3$, the models now fall within the $1\sigma$ confidence region for lower values of $\wre$. The case with $n=2/3$ lies on the edge of the $2\sigma$ contour, while the conventional $\phi^2$ model remains well outside the allowed region. Interestingly, recent studies have demonstrated that introducing a non-minimal coupling between the inflaton and gravity can shift the predictions of the $\phi^2$ model into the allowed observational region~\cite{Kallosh:2025rni}.

We now investigate whether an NBD initial state can help achieve the target spectral index $\ns=0.9743$ as indicated by the P-ACT-LB-BK18 data. To this end, we compute the required values of the NBD parameter $\beta_0$, which are presented in the middle left panel of Fig.~\ref{fig:ct-ht} as a function of the potential index $n$. In some cases, achieving the desired value of $\ns$ requires lowering it from the value obtained under the standard BD vacuum, which in turn leads to negative values of $\beta_0$. The corresponding predictions for the tensor-to-scalar ratio $r$, including the NBD corrections, are shown in the bottom left panel of Fig.~\ref{fig:ct-ht}. We observe that $r$ tends to exceed the current upper bounds in this scenario. Notably, even the case with $n=1/3$, which initially fell within the $1\sigma$ region under BD initial conditions, now predicts a larger $r$ due to the downward adjustment of $\ns$, thereby moving outside the allowed region. Therefore, for monomial potentials, the introduction of NBD initial conditions does not significantly improve consistency with the $\ns-r$ observational bounds. In such cases, it may be more promising to explore models with non-canonical kinetic terms~\cite{Fakir:1990eg,Unnikrishnan:2012zu,Unnikrishnan:2013vga,Kallosh:2025rni}.

\subsection{Hilltop inflation}

Hilltop inflation refers to a class of models where the inflaton field starts near the top of a potential hill and rolls slowly towards the minimum, enabling a prolonged period of inflation. These models are characterized by potentials with a maximum at or near the origin, typically of the form, $V=V0[1-(\phi/f)^n+\cdots]$ where $f$ sets the width of the hill and $n$ determines the steepness. Due to the flatness near the top, hilltop models can naturally produce a low tensor-to-scalar ratio 
$r$, often within observational limits.
In this work we shall consider the potential to be~\cite{Boubekeur:2005zm,Lillepalu:2022knx,Lin:2008ys,King:2024ssx,Wolf:2024lbf,Adhikari:2019uaw,Gangopadhyay:2022vgh}
\begin{align}
    V(\phi)=V_0 \l[1-\l(\f{\phi}{f}\r)^n\r],
\end{align}
and focus on the specific case with 
$n=2$. The first and second slow roll parameters for this potential are given by
\begin{align}
    \epsilon &=\f{2 \Mpl^2 \phi^2}{(f^2 - \phi^2)^2}, \quad
    \eta= -\f{2 \Mpl^2}{f^2 - \phi^2}.
\end{align}
The number of e-folds generated during inflation, under the slow roll approximation, is 
\begin{align}
\Nk=\f{1}{2}\f{f^2}{\Mpl^2}\log\l(\f{\phie}{\phi_k}\r) + \f{\phi_k^2 -\phie^2}{4\Mpl^2},
\end{align}
where, as before, $\phi_k$ and $\phie$ denote the field values at pivot scale and at the end of inflation, respectively.

In the top right panel of Fig.~\ref{fig:ct-ht}, we present the predictions for $\ns$–$r$ in hilltop inflation for two representative values of the reheating EoS parameter: $\wre = 0$ and $\wre = 0.9$. For reference, we overlay the observational contours from P-LB-BK18 and P-ACT-LB-BK18. As in previous cases, the black curves show the boundaries of the parameter space set by the maximum and minimum allowed reheating temperatures, $\Tre$. We find that the hilltop potential remains compatible with the $\ns-r$ bounds, even after the ACT data shift the preferred value of $\ns$ to higher values. To meet this updated constraint, a larger value of the parameter $f$ is needed. From the figure, we observe that the model satisfies the $1\sigma$ bound for values of $f$ in the range $12\Mpl \lesssim f \lesssim 15\Mpl$.

Although hilltop inflation can satisfy the $\ns$–$r$ constraint even with the standard BD initial state, we examine the impact of a NBD initial condition for completeness. In the middle panel of Fig.~\ref{fig:ct-ht}, we show the values of the $\beta_0$ required to achieve $\ns = 0.9743$. We observe that for a steep reheating EoS parameter $\wre = 0.9$ and for $f \gtrsim 17\Mpl$, the value of $\beta_0$ becomes negative in order to lower $\ns$ from a higher value predicted under the BD assumption. The corresponding values of $r$, incorporating the effects of the NBD correction, are displayed in the bottom right panel of Fig.~\ref{fig:ct-ht}. It is evident that the inclusion of the NBD initial state allows a broader range of the parameter $f$ to remain consistent with observational bounds on $r$.

\subsection{Natural inflation}

Natural inflation is another class of well known model where the potential takes the following form~\cite{Kim:2004rp,Yonekura:2014oja,Stein:2021uge}
\begin{align}
V(\phi)=V_0\l[1+\cos\l(\f{\phi}{f}\r)\r],
\label{eq:nat-pot}
\end{align}
where the decay constant $f$ controls the flatness of the potential. The periodic nature of the potential provides a natural mechanism to ensure the slow roll conditions, making the model technically radiatively stable.
The slow roll parameters are 
\begin{align}
    \epsilon&=\f{\Mpl^2}{2f^2}\tan^2\l(\f{\phi}{2f}\r), \\
    \eta &= -\f{\Mpl^2}{f^2} \f{\cos(\phi/f)}{1+\cos(\phi/f)},
\end{align}
and the number of inflationary e-folds
\begin{align}
    N_k=
    \f{2f^2}{\Mpl^2}\l\{\log\l[\sin\l(\f{\phie}{2f}\r)\r]-\log\l[\sin\l(\f{\phi_k}{2f}\r)\r]\r\}.
\end{align}
It is well known that natural inflation~\cite{Freese:1990rb} is under increasing tension with current cosmological observations, particularly due to its prediction of a relatively large $r$~\cite{Freese:2014nla,Reyimuaji:2020bkm,Reyimuaji:2020goi}, which is disfavored by the latest data from P-ACT-LB-BK18. Despite its appealing theoretical motivations\footnote{The flatness of the potential in natural inflation is protected by the shift symmetry, which naturally suppresses large radiative corrections, solving the $\eta$-problem. 
It is motivated by axion physics, where many string theory compactifications produce axion-like fields.
Also, such a cosine-type potential in Eq.~\eqref{eq:nat-pot} naturally arises in theories with spontaneous breaking of a global symmetry. 
The potential also sustain inflation without excessive fine-tuning. For detailed discussion regarding this, see~\cite{delaFuente:2014aca,Neupane:2014vwa,Nomura:2017ehb,Stein:2021uge}.}, 
the model struggles to remain within the favored region of the $\ns-r$ parameter space~\cite{Stein:2021uge}.
In this work, we investigate whether introducing a NBD initial state can help reconcile natural inflation with current observational bounds. Specifically, we aim to achieve the desired spectral index $\ns=0.9743$, as indicated by the P-ACT-LB-BK18 data, by tuning the parameter $\beta_0$. Our analysis reveals that in order to match the observed value of $\ns$, a relatively large value of $\beta\geq0.6$ is required. However, this modification comes at a cost. The introduction of the NBD correction significantly enhances the predicted value of the $r$, pushing it above the observational upper limit of $r\sim0.05$.
Thus, even with the inclusion of a non-standard initial  state, natural inflation remains in tension with the current data, indicating that the NBD initial condition does not resolve the fundamental issue.


\section{How large $\bk$ can be?}\label{sec:lbk}

In this section we shall discuss the viability of choosing the value of the NBD parameter $\bk$ to avoid backreaction. As already mentioned, from the gravitational backreaction, we have an upper bound on $\bk$ to be $\bk\lesssim\sqrt{\epsilon\eta'}H\Mpl/M^2$, where $M$ defines the scale of new physics~\cite{Porrati:2004gz,Greene:2004np,Ashoorioon:2017toq}. From the choice of our $\bk$ in Eq.~\eqref{eq:betak}, it is evident that the change in the power spectrum is controlled by $k/(aM)\sim H/M$. Hence maximum allowed value of $\beta_0$ can be controlled by changing the value of $H/M$. For consistency with effective field theory, $M>H$ is typically required. In this work we have chosen at pivot scale, $H_\ast/M=0.2$ as suggested in~\cite{Brahma:2025dio}. For a typical choice of $\epsilon\sim\eta'\sim 10^{-2}$ and $H\sim 10^{-5}\Mpl$ the allowed value of $\bk$ can even reach of the order of $\mathcal{O}(10)$. Which can also be understood from figure 1 of~\cite{Greene:2004np}. In such a case, our choice of $\beta_0<0.2$ lies well below the allowed value. 

There is yet another issue associated with the largeness of $\bk$ that we shall discuss. This comes from the amount of non-Gaussianity that is produced due to the NBD initial state~\cite{Martin:2000xs,Holman:2007na,Emami:2014tpa,Meerburg:2009ys,Ragavendra:2020vud}. The non-Gaussianity parameter associated to three scalar mode is given by the ratio to the three-point correlation to the power spectrum as 
\begin{align}
    \fnl&=-\f{10}{3}\f{1}{(2\pi)^4}{G_{\zeta\zeta\zeta}(\vka,\vkb,\vkc)}\\&\quad
    \times \l[\f{\ps(k_1)}{k_1^3}\f{\ps(k_2)}{k_2^3}+\text{two combinations}\r]^{-1},
\end{align}
where $G_{\zeta\zeta\zeta}(\vka,\vkb,\vkc)$ is the three-point correlation of the perturbation $\hat\zeta$, defined as 
\begin{align}
\langle\hat{\zeta}_{\vka}\hat{\zeta}_{\vkb}\hat{\zeta}_{\vkc}\rangle=\f{1}{(2\pi)^{3/2}}G_{\zeta\zeta\zeta}(\vka,\vkb,\vkc)
    \delta^{(3)}(\vka+\vkb+\vkc).
\end{align}
The observational constraints on $\fnl$ in local, equilateral and orthogonal shape is given by~\cite{Planck:2019kim}
\begin{subequations}
\begin{align}
    \fnl^{\rm local}&=-0.9\pm5.1,\\
    \fnl^{\rm equil}&=-26\pm47,\\
    \fnl^{\rm ortho}&=-38\pm24.
\end{align}
\end{subequations}
It is well known that a change in $\bk$ can significantly affect $\fnl$. Hence it becomes important to check that our choice of $\bk$ does not produce an $\fnl$ beyond the allowed range. The standard consistency condition does not hold if the initial state is not BD. The NBD state can produce non-Gaussianity which can be dominant in the flattened shape, where $\ka=\kb=\kc/2$~\cite{Holman:2007na}. The non-Gaussianity parameter in flattened shape is given by 
\begin{align}
    \fnl^{\rm flattened}\sim \f{\dot\phi^2}{M^4}\bk,
\end{align}
where after taking into account the bound from gravitational backreaction, the maximum expected value of $\fnl$ reads
\begin{align}
     \fnl^{\rm flattened}\sim \epsilon\sqrt{\epsilon\eta'}\l(\f{H}{\Mpl}\r)^2\l(\f{\Mpl}{M}\r)^5.
\end{align}
Now, taking the typical value of the parameters $\epsilon\sim\eta\sim 10^{-2}$ and  $H\sim 10^{-5}\Mpl$ we see that, the choice of $M=5H$ at pivot scale, produces large $\fnl$. One can reduce $\fnl$ down to order $\mathcal{O}(10)$, by choosing an lower value of $H/M$, say for instance,  taking $M\sim 70H$. However, adopting such a value of $M$, can exclude parts of the parameter space in $\wre-\Tre$, which might otherwise be viable for lower values of $\epsilon$, $\eta'$. A deeper analysis of specific inflationary models is required to assess these possibilities.
Hence, it is safe to say that, for some cases, $\ns$ can not be brought back to the allowed $1\sigma$ region solely by invoking a NBD initial state, without violating the observational bound on $r$, or generating an unacceptably large $\fnl$. As an alternative NBD state, one may further explore another class of states called coherent states which possibly has lesser issues due to large non-Gaussianities and backreaction~\cite{Kundu:2011sg,Kundu:2013gha,Mondal:2024glo,Ragavendra:2024qpj}.

\section{Conclusions }\label{sec:conclusions}

In this work, we have considered a class of single-field slow roll inflationary models and analysed their predictions for the scalar spectral index $\ns$ and tensor-to-scalar ratio $r$ in light of the recent combined CMB dataset from Planck 2018, ACT DR6, DESI DR1, and BICEP/Keck 2018. The inclusion of ACT data has notably shifted the preferred value of the spectral index to a higher value, $\ns = 0.9743 \pm 0.0034 $, motivating a re-examination of well-known inflationary scenarios to assess their viability under the updated observational constraints.
In particular, we have studied the predictions of the $\alpha$-attractor E- and T-models, chaotic inflation with monomial potentials, hilltop inflation, and natural inflation. Interestingly, the shift in $\ns$ now allows certain chaotic inflation models with fractional exponents, such as $n=1/3$ or $1/2$, to fall within the $1\sigma$ region, while simultaneously placing the Starobinsky model under increased tension with the data.
We have further investigated the role of the phase of reheating in modifying the inflationary predictions, including the effects of the reheating equation of state and reheating temperature. We further impose constraints from the effective number of relativistic species $\dneff$, which sets a lower bound on the reheating temperature based on the strength of primordial gravitational waves.
Additionally, we explore how deviations from the standard BD vacuum can impact the inflationary predictions. We find that, in several scenarios, the inclusion of NBD initial conditions can further improve the agreement of inflationary models with observational constraints, even in cases that were previously excluded after accounting for reheating dynamics.
We now proceed to summarise the key findings of this study.

In the case of the $\alpha$-attractor E- and T-models, we find that introducing a steep EoS during the reheating phase can shift the predicted value of $\ns$ closer to the observed central value, allowing the model to enter the $2\sigma$ confidence region. However, even with this modification, the predictions typically fall short of reaching the $1\sigma$ region. Remarkably, the inclusion of a NBD initial state, with a small but non-zero value of the parameter $\beta_0$ can raise $\ns$ sufficiently to match the preferred observational value of $\ns = 0.9743$. Additionally, the corresponding modified $r$ remains well within the observational upper bound of $r<0.036$ across a broad range of the model parameter $\alpha$, further enhancing the consistency of the $\alpha$-attractor scenario with current data.

For chaotic inflation, we observe that with a low reheating equation of state parameter $\wre$, monomial potentials of the form $\phi^{1/5}$ and $\phi^{1/3}$ can satisfy the $1\sigma$ bounds on $\ns$ and $r$, while the $\phi^{1/2}$ potential lies within the $2\sigma $ region. However, the canonical slow roll scenario with a quadratic potential  $m^2 \phi^2$ remains strongly disfavored by the data. Upon introducing a NBD initial state, we find that the required value of $\beta_0$ to achieve $\ns=0.9743$ can be either significantly positive or negative, depending on the model parameters. Despite this, the incorporation of NBD initial conditions does not substantially improve the compatibility of chaotic inflation with observations. In particular, for the $\phi^{1/3}$ potential, which was previously consistent with the data once reheating effects were included, the NBD correction tends to increase $r$  beyond the observational upper limit. This suggests that in order to reconcile chaotic inflation with current constraints, one may need to consider extensions involving non-canonical kinetic terms or modified gravity frameworks.

Hilltop inflation proves to be more promising in light of the updated observational constraints by ACT. Despite the shift of the $\ns-r$ contour towards higher values of $\ns$, the hilltop potential remains consistent with the $1\sigma$ bound for relatively large values of the parameter $f$, specifically in the range $12\Mpl \lesssim f \lesssim 15\Mpl$. To explore possible improvements, we also examine the impact of a NBD initial state. We find that introducing the NBD correction allows a broader range of $f$, including values below $12\Mpl$, to achieve the mean value of $\ns$ while keeping $r$ within the observationally allowed limit. 

Lastly, natural inflation remains in significant tension with the current observational data, even after incorporating a phase of reheating or introducing a NBD initial condition. A similar conclusion holds for chaotic inflation with higher powers of the potential, such as the quadratic case. These findings indicate that modest extensions using NBD initial state within the canonical framework are insufficient to reconcile these models with the data. Therefore, more substantial modifications, such as introducing non-canonical kinetic terms, exploring multi-field inflation, may be necessary for these classes of inflationary models to remain observationally viable.

\section*{Acknowledgments}

SM thanks the Indian Institute of Science Education and Research (IISER), Pune for support through a postdoctoral fellowship. SM also wishes to thank Arka Banerjee, Diptimoy Ghosh, Susmita Adhikari, Sachin Jain, and Farman Ullah for useful discussions. SM would also like to thank Shahin Sheikh Jabbari, Amjad Ashoorioon, and H.V. Ragavendra for their valuable comments.


\bibliographystyle{apsrev4-1}
\bibliography{references}

\end{document}